\documentclass[letter,twocolumn]{jpsj2}

\def\runauthor{Kazuaki {\sc Iwasa} {\it et al.}}

\title
{
Role of $p$-$f$ Hybridization in the Metal-Non-Metal Transition of PrRu$_4$P$_{12}$
}

\author
{ 
Kazuaki {\sc Iwasa}\thanks{E-mail: iwasa@iiyo.phys.tohoku.ac.jp},
Lijie {\sc Hao},
Tomoo {\sc Hasegawa},
Toshiaki {\sc Takagi},
Kenji {\sc Horiuchi},\\
Yoshiaki {\sc Mori},
Youichi {\sc Murakami},
Keitaro {\sc Kuwahara}$^{1}$,
Masahumi {\sc Kohgi}$^{1}$,\\
Hitoshi {\sc Sugawara}$^{2}$,
Shanta Ranjan {\sc Saha}$^{3}$\thanks{Present address: Department of Physics \& Astronomy, Faculty of Science, McMaster University, Canada.},
Yuji {\sc Aoki}$^{1}$
and
Hideyuki {\sc Sato}$^{1}$
}

\inst
{
Department of Physics, Tohoku University, Sendai 980-8578\\
$^1$Department of Physics,Tokyo Metropolitan University, Hachioji, Tokyo 192-0397\\
$^2$Department of Mathematical and Natural Sciences, 
The University of Tokushima, Tokushima 770-8502\\
$^3$Muon Science Laboratory, Institute of Material Structure Science, High Energy Accelerator Research Organization, Tsukuba, Ibaraki 305-0801\\
}

\recdate
{
\today
}

\abst
{
Electronic state evolution in the metal-non-metal transition of PrRu$_4$P$_{12}$ has been studied by X-ray and polarized neutron diffraction experiments. It has been revealed that, in the low-temperature non-metallic phase, two inequivalent crystal-field (CF) schemes of Pr$^{3+}$ $4f^2$ electrons with $\Gamma_1$ and $\Gamma_4^{(2)}$ ground states are located at Pr1 and Pr2 sites forming the bcc unit cell surrounded by the smaller and larger cubic Ru-ion sublattices, respectively. This modulated electronic state can be explained by the $p$-$f$ hybridization mechanism taking two intermediate states of $4f^1$ and $4f^3$. The $p$-$f$ hybridization effect plays an important role for the electronic energy gain in the metal-non-metal transition originated from the Fermi surface nesting.
}

\kword
{
metal-non-metal transition, Fermi surface nesting, crystal-field state, $p$-$f$ hybridization
}

\begin{document}
\maketitle

Various electronic properties of rare-earth-filled skutterudites RT$_4$X$_{12}$ (R = rare earth, T = transition metal, X = pnictogen) are recent topics in the field of strongly correlated electron systems~\cite{Aoki_JPSJKondo}. For example, some Pr-based compounds exhibit heavy-electron states and quadrupolar orderings. It is remarkable that hybridization between R-ion $4f$- and X-atom $p$-electron states is crucial for these various properties because of the characteristic crystal structure where a R ion is surrounded by twelve X atoms~\cite{Harima03}. 

PrRu$_4$P$_{12}$, which is the subject material in the present study, has been reported to show a metal-insulator (M-I) transition at $T_{\rm M-I} \simeq 63$ K~\cite{Sekine97}. Electron diffraction experiments elucidated the structural transformation from a body-centered cubic crystal structure (space group Im${\bar 3}$) in the high-temperature metallic phase to a simple cubic one in the low-temperature insulator phase~\cite{Lee01}. The phase transition has been interpreted as a formation of charge density wave (CDW) due to the nesting condition of the Fermi surface, as proposed in the band calculation study~\cite{Harima02}. However, the electrical resistivity has a shoulder-like anomaly around 40 K and stays at a finite value at the measured lowest temperature. Large negative magnetoresistance was also observed under magnetic field less than only 0.5 T at 0.45 K~\cite{Saha_R_03}. These facts indicate an incomplete insulator phase below $T_{\rm M-I}$, which is different from the expected simple CDW formation. Since LaRu$_4$P$_{12}$ does not undergo transition to non-metallic phase in spite of its Fermi surface similar to that of PrRu$_4$P$_{12}$~\cite {Harima02,Saha_dHvA}, it is naturally expected that  the Pr-ion 4f electrons play an important role in the metal-nonmetal transition. In order to investigate this interesting issue, we have carried out inelastic neutron scattering experiments~\cite{Iwasa_inela}. Above $T_{\rm M-I}$, there are overdamped crystal-field (CF) excitations which are attributed to the strong $p$-$f$ hybridization, and the ground state was assigned to a non-magnetic singlet $\Gamma_1$. With decreasing temperature below $T_{\rm M-I}$, the excitation peaks shift by a few meV and become sharper, indicating that the $4f$-electron state becomes localized with lowering temperature. We have concluded that the hybridization effect enhances the density of states at the Fermi level and causes the metal-non-metal transition of PrRu$_4$P$_{12}$. It should be noted that the Pr-ion sites split into two inequivalent CF-level schemes below $T_{\rm M-I}$. One of them has a ground state $\Gamma_1$ and the other switches to a magnetic triplet $\Gamma_4^{(2)}$ at the lowest temperature. Such two inequivalent CF level schemes reproduce well the bulk magnetic behavior~\cite{Sekine97,Iwasa_inela}. 

The metal-non-metal transition of PrRu$_4$P$_{12}$ accompanied by the drastic thermal evolution of the CF schemes is a new phenomenon owing to the electron correlation involving the $4f$ states. The purpose of the present study is to clarify its mechanism by determining the mutual arrangement between the atomic displacements and the CF schemes and by comparing the experimental result with the recent theoretical proposition for the $p$-$f$ hybridization effect in the Pr-filled skutterudite compounds~\cite{Otsuki_CF}. To accomplish this purpose, we have performed X-ray and polarized neutron diffraction experiments. A part of the X-ray diffraction study has been published elsewhere~\cite{Hao04}.

Polyhedron-shaped single-crystal samples of PrRu$_4$P$_{12}$ around 2 mm in diameter were synthesized by Sn-flux method as adopted in the transport study~\cite{Saha_R_03}. X-ray diffraction experiments were performed using a rotating-anode (Mo) X-ray generator and a diffractometer equipped with a lifting detector. The sample was cooled down to 7 K by a helium-gas closed-cycle refrigerator. Neutron diffraction experiment was performed at the thermal neutron spectrometers PONTA installed at the beam hole 5G and TOPAN at 6G of JRR-3M reactor in JAERI, Tokai, Japan. Spin polarized neutron beam of wave number 2.55 ${\rm \AA}^{-1}$ is selected by a Heusler alloy monochromator. A single-crystal sample set in a cryomagnet was cooled between 2.2 and 20 K, and magnetic fields up to 4 T were applied vertically along the crystalline axis [0~0~1]. We measured diffraction intensities for two incident neutron spin directions, parallel or anti-parallel to the induced magnetic moments, switched by an electromagnetic flipper. The intensity at the reciprocal-lattice point $\textbf{Q}$ of PrRu$_4$P$_{12}$ is expressed as $I = | F_{\rm N}(\textbf{Q}) | ^2 + | F_{\rm M}(\textbf{Q}) | ^2 + 2$\textbf{P}{\boldmath $\cdot  \bar{\mu}$}$F_{\rm N}(\textbf{Q})F_{\rm M}(\textbf{Q})$, where $F_{\rm N}({\bf Q})$, $F_{\rm M}(\textbf{Q})$, $\textbf{P}$ and \mbox{\boldmath $\bar{\mu}$} are a nuclear structure factor, a magnetic structure factor, the neutron spin polarization and a unit vector parallel to the induced magnetic moments, respectively. Hereafter, the intensity $I_+$ is defined for the case of \textbf{P}{\boldmath $\cdot  \bar{\mu}$ = +1}, and $I_-$ for \textbf{P}{\boldmath $\cdot  \bar{\mu}$} $= -1$.

\begin{figure}[t]
\begin{center}
\includegraphics[width=8cm]{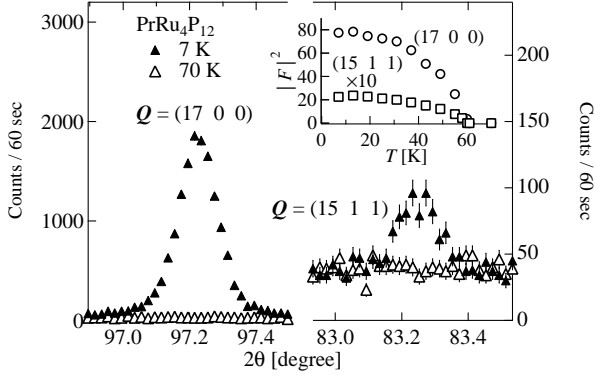}
\end{center}
\caption{X-ray superlattice reflection profiles at $\textbf{Q}  = (17~0~0)$ and $\textbf{Q}  = (15~1~1)$. An inset shows temperature dependencies of squared structure factors for each reflections.}
\label{f1}
\end{figure}

\begin{figure}[t]
\begin{center}
\includegraphics[width=8cm]{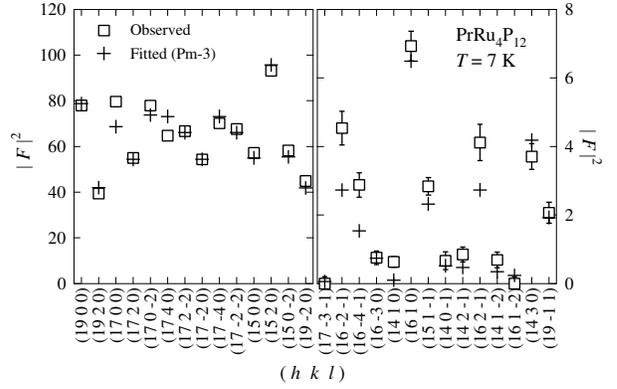}
\end{center}
\caption{Squared structure factors of the X-ray superlattice reflections at 7 K. Open squares are experimentally determined values, and cross symbols fitted results based on the Pm${\bar 3}$ model.}
\label{f2}
\end{figure}

Figure~\ref{f1} shows X-ray diffraction profiles at $\textbf{Q} = (17~0~0)$ and $(15~1~1)$  corresponding to the prohibited positions of the high-temperature structure. 
The inset depicts temperature dependencies of their squared structure factors, which were estimated from the measure intensities of the neighboring fundamental reflections. Since the reflections grow up below $T_{\rm M-I}$, the low-temperature phase is characterized by the superlattice formation represented by the modulation wave vector $\textbf{q} = (1~0~0)$. We have obtained structure factors of various superlattice reflections at 7 K, as shown in Fig.~\ref{f2} by open squares. 

Possible atomic displacements in the low temperature phase of PrRu$_4$P$_{12}$ have been proposed on the basis of a symmetry argument~\cite{Curnoe}. Since there is no evidence for any $4f$-electron magnetic or multipolar ordering down to 0.15 K~\cite{Saha_R_03}, the local symmetry of Pr-ion sites was considered to be conserved. Thus the low-temperature structure was assumed to be a simple cubic of space groups Pm${\bar 3}$ or P23. The former model was adopted in the band calculation, and an electronic gap formation was obtained by taking into account P-atom displacements~\cite{Harima03_PrRuP}. The atomic coordinates in the Pm${\bar 3}$ cubic unit cell are represented by $(0, u + \delta_u, v + \delta_v)$ for 12j site and $(1/2, 1/2 + u - \delta_u, 1/2 + v - \delta_v)$ for 12k site, where $u = 0.3576$ and $v = 0.1444$ are in the high-temperature phase~\cite{Harima03_PrRuP}. The Ru-ion displacement was also taken into account, and its atomic coordinate of 8i site is expressed as $(1/4+\delta, 1/4+\delta, 1/4+\delta)$. We performed a least-squares fitting of the calculated structure factors with free parameters of $\delta$, $\delta_u$ and $\delta_v$ to the present experimental values. The result shown by cross symbols in Fig.~\ref{f2} reproduces the data satisfactorily. The best fit parameters are $\delta = 7 \times 10^{-4}$, $\delta_u = -3 \times 10^{-4}$ and $\delta_v = 6 \times 10^{-4}$, which are very close to those by another X-ray diffraction experiment~\cite{Lee04}. The model of P23 structure was also examined, and it also reproduces the experimental data similarly well. The resultant atomic coordinates are close to those of the Pm${\bar 3}$ model, despite that there are twice independent positions at the Ru-ion 4e sites compared to the Pm${\bar 3}$ case~\cite{Hao04,Lee04}. Thus, we can conclude that the Pm${\bar 3}$ model is sufficient for expressing the superlattice structure. The P-atom displacement is one order of magnitude smaller than that in the band calculation result~\cite{Harima03_PrRuP}. The Ru-ion displacement is necessary to reproduce the experimental data satisfactorily. It is remarkable that the Ru ions move toward one of the neighboring Pr ions at the unit-cell center or at the corner and form two different cubic configurations around the Pr ions, although the size of P-atom cage surrounding the Pr ions does not change remarkably. Because of the structural superlattice, one can expect that the inequivalent Pr ions in the unit cell take the different CF schemes.

\begin{figure}[t]
\begin{center}
\includegraphics[width=8cm]{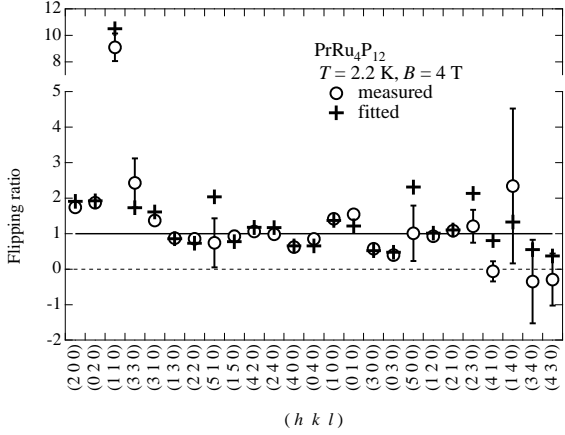}
\end{center}
\caption{Circles are flipping ratios $I_+ / I_-$ for various reflection points measured at 2.2 K and at 4 T. Crosses are fitted results.}
\label{f3}
\end{figure}

\begin{figure}[t]
\begin{center}
\includegraphics[width=8cm]{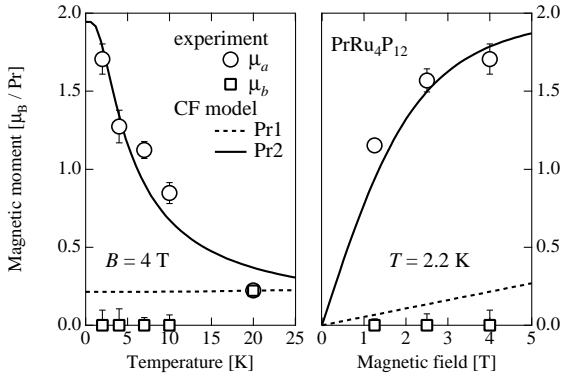}
\end{center}
\caption{Symbols are experimentally determined field-induced magnetic moments $\mu_a$ at the unit-cell corner site and $\mu_b$ at the center site as functions of temperature and magnetic field.  Broken lines are calculated results for the CF scheme at Pr1 ($\Gamma_1$ ground state), and solid lines for Pr2 ($\Gamma_4^{(2)}$ ground state).}
\label{f4}
\end{figure}

Next, we present the polarized neutron diffraction results to show the arrangement of the two inequivalent Pr-ion CF schemes in the structural superlattice. Circles in Fig.~\ref{f3} shows the flipping ratio $I_+ / I_-$ at 2.2 K and at 4 T for various $\textbf{Q}$ positions not only for the superlattice but also for the fundamental reflections. 
It is obvious that the many flipping ratios deviates from unity, indicating an interference between the reflection amplitudes of $F_{\rm N}(\textbf{Q})$ and $F_{\rm M}(\textbf{Q})$ even for the superlattice peaks. Thus, there is an antiferromagnetic component characterized by $\textbf{q} = (1~0~0)$ which is the same as that of the structural superlattice. Although the data are not shown here, $I_+ / I_-$ become closer to unity as the temperature is increased or the applied magnetic field is decreased. We have carried out a least-squares refinement of $I_+ / I_-$ using the two induced magnetic-moment magnitudes $\mu_a$ and $\mu_b$ at the unit-cell center and the corner, respectively, as free parameters. The magnetic form factor of Pr$^{3+}$ ion is quoted from the calculation within the dipole approximation~\cite{Pr_magff}. The atomic coordinates were fixed to the aforementioned values at 7 K, because the atomic displacement is almost independent of temperature below 20 K as seen in the temperature dependence of the superlattice structure factors shown in the inset of Fig.~\ref{f1}. Moreover, $T_{\rm M-I}$ is independent of magnetic field~\cite{Sekine_C}, so that it is naturally expected that the atomic displacements also show no distinct magnetic-field dependence. The fitted result depicted by cross symbols in Fig.~\ref{f3} agrees quite well with the observation. The resultant induced magnetic moments are $\mu_a = (1.7 \pm 0.1) \mu_{\rm B}$ and $\mu_b = (0 \pm 0.1) \mu_{\rm B}$. It is noticeable that the Pr ion at the unit-cell center to which the Ru ions moves closer shows a smaller magnetic moment. Symbols in Fig.~\ref{f4} depict the best fitted values of $\mu_a$ and $\mu_b$ as functions of temperature and applied magnetic field.

\begin{figure}[t]
\begin{center}
\includegraphics[width=8cm]{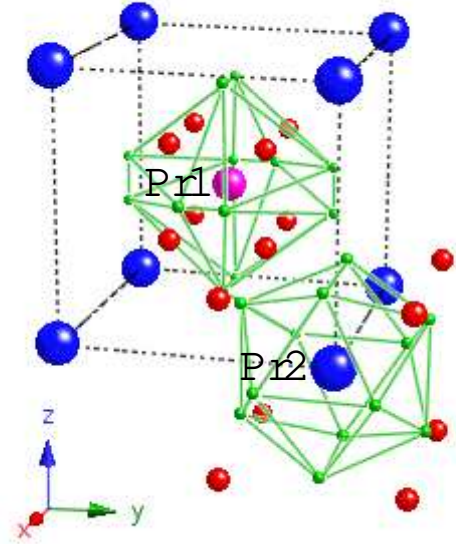}
\end{center}
\caption{Schematic drawing of the superlattice in the non-metallic phase of PrRu$_4$P$_{12}$. Pr1 and Pr2 indicate the Pr ions with the ground state $\Gamma_1$ and $\Gamma_4^{(2)}$, respectively. Atomic displacements of Ru and P are enlarged by a hundred times.}
\label{f5}
\end{figure}

This result can be compared to the CF schemes determined in the polycrystalline inelastic neutron scattering study~\cite{Iwasa_inela}. Because they hardly change below 20 K, we take $\Gamma_1 {\rm (0~meV)} - \Gamma_4^{(2)} {\rm (8.07~meV)} - \Gamma_4^{(1)} {\rm (9.18~meV)} - \Gamma_{23} {\rm (15.71~meV)}$ for Pr1 and $\Gamma_4^{(2)} {\rm (0~meV)} - \Gamma_1 {\rm (3.12~meV)} - \Gamma_4^{(1)} {\rm (13.88~meV)} - \Gamma_{23} {\rm (20.13~meV)}$ for Pr2 at 5 K. The induced magnetic moments at Pr1 and Pr2, calculated by taking into account the Zeeman term as well as the CF Hamiltonian, are shown by broken and solid lines in Fig.~\ref{f4}, respectively. The CF schemes of Pr1 and  for Pr2 explain reasonably the experimentally obtained $\mu_b$ and $\mu_a$, respectively.~\cite{note1} Consequently, the larger magnetic moment from the Pr2 scheme at the corner site and the smaller one from Pr1 at the center site are necessary to reproduce the overall feature of observed $I_+ / I_-$. The resultant low-temperature superlattice of the atomic displacements and the Pr-ion CF schemes is depicted schematically in Fig.~\ref{f5}, where the atomic displacements are enlarged by a hundred times.

The CF level scheme for Pr$^{3+}$ $4f^2$ state in Pr-filled skutterudites was recently studied by Otsuki {\it et al}.~\cite{Otsuki_CF}. In the case of only the point charge Coulomb potential, the scheme $\Gamma_1 - \Gamma_4^{(1)} - \Gamma_{23} - \Gamma_4^{(2)}$ is favored under the condition of positive valence of transition-metal ions. It can be modified by the hybridization of $4f$ states with the $a_u$ conduction band originated dominantly from pnictogen $p$ orbits in the filled skutterudites~\cite{Harima03}. This $p$-$f$ hybridization effect was treated as the perturbation involving two channels of intermediate states of $4f^1$ with creation of an electron in the vacant states and $4f^3$ with a hole in filled states. It is demonstrated that the $4f^3$ process pulls down the $\Gamma_4^{(2)}$ triplet, and the $4f^1$ process works in an opposite way. The experimentally determined CF scheme at 70 K above $T_{\rm M-I}$ is $\Gamma_1 {\rm (0~meV)} - \Gamma_4^{(1)} {\rm (5.87~meV)} - \Gamma_4^{(2)} {\rm (10.83~meV)} - \Gamma_{23} {\rm (13.27~meV)}$. The reversed sequence of $\Gamma_4^{(2)}$ and $\Gamma_{23}$ from the point-charge scheme indicates the significant hybridization effect, which is consistent with the observed broad CF excitation spectra~\cite{Iwasa_inela}. However, the contributions of both the $4f^1$ and $4f^3$ processes are comparable, so that the total effect on the CF splits may be not enough to pull down $\Gamma_4^{(2)}$ to much lower energy level. With decreasing temperature below $T_{\rm M-I}$, the Fermi surface starts to vanish and the $4f$ electrons shift to rather localized state as seen in the sharp CF excitation spectra. The CF schemes at Pr1 and Pr2 switch to those described in the preceding paragraph. The $\Gamma_4^{(2)}$ states at both Pr ions shifting to lower levels than that around $T_{\rm M-I}$ can be ascribed to the $p$-$f$ hybridization with relatively larger effect of the $4f^3$ process than that of $4f^1$. From the incomplete insulator nature observed at the lowest temperature~\cite{Sekine97,Saha_R_03}, we can expect that some part of the Fermi surface survives below $T_{\rm M-I}$. Hereafter, we assume a semi-metallic band structure for PrRu$_4$P$_{12}$ at low temperatures, although it has not been established for this compound so far and the resisitivity shows upturn below $T_{\rm M-I}$. The resistivity upturn may come from low mobility of the low-density carrier of the semi-metallic band owing to scattering by impurities and defects in the crystal~\cite{note2}. The hybridization between the valence band and $f$ electrons lowers the $4f$ level and raises the valence band as a result of the bonding-antibonding effect~\cite{Kasuya}. It will give rise to the increase of the hole number and the shift of Fermi energy due to re-arrangement of electrons to the conduction bands. This process can give actual energy gain~\cite{Kasuya} and is consistent with the non-divergent resistivity below $T_{\rm M-I}$. Thus, the $p$-$f$ hybridization can be responsible not only for the CF splits but also for the stability of the incomplete insulator state below $T_{\rm M-I}$. The result of the analysis of the inelastic neutron scattering experiment shows that the $\Gamma_4^{(2)}$ state at Pr2 shifts drastically to the ground state below $T_{\rm M-I}$, while the low-temperature CF scheme at Pr1 is similar to that around $T_{\rm M-I}$. This different behaviors between the two Pr sites can be interpreted as that the contribution of the $4f^3$ process is larger than the $4f^1$ process at Pr2 and the contributions of both processes are comparable at Pr1. The appearance of the two different 4$f$-electron states of Pr ions in the low temperature phase of PrRu$_4$P$_{12}$ is naturally understood as a resullt of the $p$-$f$ hybridization effect combined with the formation of charge density modulation with the wave vector $\textbf{q} = (1~0~0)$ expected from the almost perfect Fermi-surface nesting. Due to this mechanism, PrRu$_4$P$_{12}$ undergoes the metal-non-metal transition in contrast to LaRu$_4$P$_{12}$.
In addition to the $p$-$f$ hybridization effect, the atomic displacement also may affect on the CF split. The positive-valence Ru-ion displacement closer to Pr1 and further from Pr2 results in the larger and smaller point-charge Coulomb potential for the Pr $4f$ electrons. It is consistent with the observed different CF schemes at Pr1 ($\Gamma_1$ ground state and $\Gamma_4^{(2)}$ excited level) and Pr2 ($\Gamma_4^{(2)}$ ground state and $\Gamma_1$ excited level). However, the atomic displacement of the order of 10$^{-3}$ is too small to dominate the evolution of the CF schemes. Thus, the $p$-$f$ hybridization plays a main role in the 4$f$-state variation in the low-temperature region. 

We have determined the mutual arrangement of the two CF schemes with the atomic displacements appearing in the metal-non-metal transition of PrRu$_4$P$_{12}$. It is demonstrated that the $p$-$f$ hybridization mechanism contributes to the electronic band energy gain resulting in the superlattice formation under the Fermi-surface nesting condition. Thus, the $p$-$f$ hybridization can play an important role in the novel metal-non-metal transition of this material.

\section*{Acknowledgements}
The authors greatly acknowledge Prof.~H.~Harima, Dr.~K.~Matsuhira, Dr.~C.~H.~Lee, Prof.~Y.~Kuramoto and Mr.~J.~Otsuki for their fruitful discussions. A part of this study is supported by the Grants-in-Aid for Scientific Research from MEXT of Japan (Young Scientists (B) (No.~15740219) and Scientific Research Priority Area ``Skutterudite" (No.~15072206)). One of the authors (L.~H.) is supported by the JSPS Postdoctoral Fellowship for Foreign Researchers. The neutron diffraction experiments were performed under the User Program conducted by ISSP, University of Tokyo.


\begin{thebibliography}{99}
\bibitem{note1} The discrepancy between the calculation and the experimental results comes from  a worse accuracy of $| \mu_a$ - $\mu_b |$ by the less number of measured antiferromagnetic reflections compared to the ferromagnetic ones giving the averaged moment value. In addition, the used magnetic form factor may not appropriate for the strongly hybridized $4f$ electrons.
\end{thebibliography}
\end{document}